\documentclass{ws-ijmpd}

\begin{document}

\markboth{Mariana P. Lima, Sandro D.P. Vitenti, Marcelo J. Rebou\c{c}as}
{Revisiting the Confrontation of the energy conditions bounds with supernovae data}

%
\catchline{}{}{}{}{}
%

\title{REVISITING THE CONFRONTATION OF THE ENERGY CONDITIONS WITH 
SUPERNOVAE DATA}

\author{MARIANA P. LIMA, SANDRO D. P. VITENTI and MARCELO J. REBOU\c{C}AS}

\address{Centro Brasileiro de Pesquisas F\'{\i}sicas \\ 
Rua Dr. Xavier Sigaud 150\\
22290-180 \ Rio de Janeiro -- RJ, Brazil 
\\
penna@cbpf.br, vitenti@cbpf.br, reboucas@cbpf.br}

\maketitle

\begin{history}
\received{Day Month Year}
\revised{Day Month Year}
\comby{Managing Editor}
\end{history}

\begin{abstract}
In the standard Friedmann--Lema\^{\i}tre--Robertson--Walker (FLRW) 
approach to model the Universe the violation of the so-called energy 
conditions is related to some important properties of the Universe 
as, for example, the current and the inflationary accelerating 
expansion phases. The energy conditions are also necessary in 
the formulation and proofs of Hawking-Penrose singularity theorems. 
In two recent articles  we have derived bounds from energy 
conditions and made confrontations of these bounds with 
supernovae data. Here, we extend these results in following
way: first, by using our most recent statistical procedure for 
calculating new $q(z)$ estimates from the \emph{gold} and
\emph{combined} type Ia supernovae samples; second, we
use these estimates to obtain a new picture of the energy 
conditions fulfillment and violation for the recent past 
($z\leq 1 $) in the context of the standard cosmology. 
\end{abstract}


\section{Introduction} 

In the absence of constraints on the energy-momentum tensor $T_{\mu\nu}$
any metric satisfies Einstein's equations since they can be regarded as a 
definition of $T_{\mu\nu}\,$, i.e., a set of equations determining 
$T_{\mu\nu}$ for any given metric $g_{\mu \nu}\,$. 
However, if one wishes to explore general properties that hold for 
a variety of different physical sources it is convenient to impose the 
so-called \emph{energy conditions} that limit the arbitrariness of 
$T_{\mu\nu}$ on physical grounds.\cite{EC-basics_refs} 

On scales relevant for cosmology, an important point in the  study of the 
energy conditions is the confrontation of their predictions with 
the observational data. By using model-independent energy-condition 
\emph{integrated} bounds on the cosmological observables as, for example,  
the distance modulus and lookback time, this confrontation has been made
in some recent articles\cite{Santos2006}\cdash\cite{CattoenVisser} (see 
also the pioneering Refs.~\refcite{M_Visser1997}  by  Visser).
In Ref.~\refcite{Lima2008a}, however,  it was shown that the fulfillment
(or the violation) of these \emph{integrated} bounds at a specific redshift 
$z$ is not a sufficient (nor a necessary) \emph{local} condition for the 
fulfillment (or respectively the violation) of the associated energy 
condition at $z$.%
\footnote{Energy conditions constraints on the so-called $f(R)$--gravity 
have also been investigated in Ref.~\refcite{Santiago2006} and more recently 
in Refs.~\refcite{SARC2007} and \refcite{SRA2008}.}
In this way, the confrontation between the prediction of these \emph{integrated} 
bounds and observational data cannot be used to draw conclusions on the 
fulfillment (or violation) of the energy conditions at $z$. 
In Ref.~\refcite{Lima2008a} this problem was overcome by deriving
new \emph{non-integrated} energy-condition bounds, and  confrontations 
between the new bounds with type Ia supernovae (SNe Ia) data of 
the \emph{gold}\cite{Riess2007} and  \emph{combined}\cite{Combined} 
samples were performed by using the  upper and lower limits of confidence 
regions on $\,E(z) - q(z)\,$ plane.  
More recently, in Ref.~\refcite{Lima2008b} a new statistical way 
for estimating the deceleration parameter $\,q(z)\,$ was carried out, 
and a new picture of the energy conditions fulfillment and violation 
for recent past ($z\leq1 $) was calculated by using the recently 
compiled \emph{Union} sample.\cite{Kowalski2008} 

In this work, we use the most recent statistical procedure introduced in 
Ref.~\refcite{Lima2008b} along with the \emph{gold}\cite{Riess2007} and  \emph{combined}\cite{Combined} samples to obtain estimates of $q(z)$ in 
order to build up a new picture of the confrontation between the energy 
condition \emph{integrated} bounds and these SNe Ia data sets, completing 
therefore the cycle of this type of analysis which involves these three 
samples and the  two statistical procedures to $q(z)$ estimates of 
Refs.\refcite{Lima2008a} and \refcite{Lima2008b}.

\section{Preliminaries}
\label{Sec2}

It is known that the energy conditions can be stated in a coordinate-invariant 
way in terms of $T_{\mu\nu}$ and vector fields of fixed character (timelike, 
null and spacelike). However, within the framework of the standard Friedmann-Lema\^{\i}tre-Robertson-Walker (FLRW) model, we only need to 
consider the energy-momentum tensor of a perfect fluid with density $\rho$ 
and pressure $p\,$, i.e., 
$T_{\mu\nu} = (\rho+p)\,u_\mu u_\nu - p \,g_{\mu \nu}\,$.
In this context, the energy conditions take the following 
forms:\cite{EC-basics_refs}
\begin{equation} \label{ec}  
\begin{array}{lll}
\mbox{NEC}: \  &\, \rho + p \geq 0 \;,  &   \\
\\
\mbox{WEC}: \ \  & \rho \geq 0 &
\ \mbox{and} \quad\, \rho + p \geq 0 \;,  \\
\\
\mbox{SEC}:   & \rho + 3p \geq 0 &
\ \mbox{and} \quad\, \rho + p \geq 0 \;, \\
\\
\mbox{DEC}:    & \rho \geq 0  &
\ \mbox{and} \; -\rho \leq p \leq\rho \;,
\end{array}
\end{equation}
where NEC, WEC,  SEC and DEC correspond, respectively, to the null, weak, 
strong, and dominant energy conditions.  For a FLRW metric with a 
scale factor $a(t)$, the density $\rho$ and pressure
$p$ of the cosmological fluid are given by 
\begin{equation}
\label{rho-p-eq}
\rho  =  \frac{3}{8\pi G}\left[\,\frac{\dot{a}^2}{a^2}
                                  +\frac{k}{a^2} \,\right]
\qquad  \text{and} \qquad
p =  - \frac{1}{8\pi G}\left[\, 2\,\frac{\ddot{a}}{a} +
\frac{\dot{a}^2}{a^2} + \frac{k}{a^2} \,\right] \;,
\end{equation}
where overdots denote the derivative with respect to the 
time $t$ and $G$ is Newton's gravitational constant.

The \emph{non-integrated} bounds from energy conditions
derived in Ref.~\refcite{Lima2008a} can be obtained in terms 
of the deceleration parameter $q(z) = -\ddot{a}/{aH^2}$, the 
normalized Hubble function $E(z) = H(z)/H_0\,$, and the curvature 
density  parameter 
$\Omega_{k0} = - k/(a_0H_0)^2$, simply by substituting 
Eqs.~\eqref{rho-p-eq} into Eqs.~\eqref{ec}. This gives%
\footnote{Through out this paper we use the notation of
Ref.~\refcite{Lima2008a} in which \textbf{NEC}, \textbf{WEC}, 
\textbf{SEC} and \textbf{DEC} correspond, respectively, 
to $\rho + p \geq 0$, $\rho \geq 0$, $\rho + 3p \geq 0$ and 
$\rho - p \geq 0$.} 
\begin{eqnarray}
\label{eq:nec-q(z)}
\mbox{\bf NEC} & \,\Leftrightarrow & \;\, q(z) - \Omega_{k0} 
\frac{(1+z)^2}{E^2(z)}   \,\geq -1 \;, \\
\label{eq:wec-omega} 
\mbox{\bf WEC} & \, \Leftrightarrow & \;\, \frac{E^2(z)}{(1 + z)^2} 
\,\geq \Omega_{k0} \;, \\
\label{eq:sec-q(z)} 
\mbox{\bf SEC} & \, \Leftrightarrow & \;\, q(z) \,\geq 0 \;, \\
\label{eq:dec-q(z)}  
\mbox{\bf DEC} & \, \Leftrightarrow & \;\, q(z) + 2\,\Omega_{k0}
\,\frac{(1+z)^2 }{E^2(z)} \,\leq  2 \;,
\end{eqnarray}
where $z = (a_0/a) -1$ is the redshift, $H(z) = \dot{a}/a\,$, 
and the subscript 0 stands for present-day quantities. 

In this work, we focus on the FLRW flat ($\Omega_{k0}= 0$) universe.
In this case the \textbf{NEC}, \textbf{SEC} and  \textbf{DEC} bounds 
reduce, respectively, to $q(z) \geq -1\,$, $q(z) \geq 0\,$ and 
$q(z) \leq 2$, while the \textbf{WEC} bound is fulfilled identically.
Thus, having estimates of $q(z_\star)$ for different redshifts $z_\star$, 
one can test the fulfillment or violation of the energy conditions 
at each $z_\star\,$. 

Now, the $q(z_\star)$ estimates are obtained by using a SNe Ia data set, 
by approximating the deceleration parameter $q(z)$ function as 
the following linear piecewise continuous function:  
\begin{equation}
\label{eq:q_z}
q(z) = q_l + q^\prime_l \, \Delta z_l \;, \quad z \in (z_l, z_{l+1})\;,
\end{equation}
where the subscript $l$ means that the quantity is taken at $z_l\,$, 
$\Delta z_l \equiv (z-z_l)\,$, and the prime denotes the derivative 
with respect to $z$. 
The supernovae observations provide the redshifts  and distance 
modulus 
\begin{equation}
\label{modulus}
\mu(z) = 5 \,\log_{10} \left[ \frac{c\, (1+z)}{H_0 \,1 \text{Mpc}} 
\;
\int_0^z \frac{\mathrm{d}z^\prime}
{E(z^\prime)}\right] + 25 \;.
\end{equation}
Then, by using the following well known relation between $q(z)$ and $E(z)$: 
\begin{equation} \label{eq:E_z}
E(z) = \exp{\int_0^z \frac{1+q(z')}{1+z'}\,\,\mathrm{d}z'} \;,
\end{equation}
along with  Eq.~\eqref{modulus}, we fitted the parameters of the $q(z)$,
as given by \eqref{eq:q_z}, by using the SNe Ia redshift--distance 
modulus data from the \emph{gold}\cite{Riess2007} and  
\emph{combined}\cite{Combined} samples.

\section{Results and Conclusions}

Since in the flat case the energy condition bounds given by 
Eqs.~\eqref{eq:nec-q(z)}, \eqref{eq:sec-q(z)} and \eqref{eq:dec-q(z)} 
depend only on $q(z)$, we have obtained the $q(z_{\star})$ estimates 
at $1\sigma - 3\sigma$ confidence levels from \emph{gold} and 
\emph{combined} SNe Ia samples by marginalizing over $E(z_\star)$ 
and the other parameters ($q'_l$ 's) of the $q(z)$ function 
[Eq.(\ref{eq:q_z})].%
\footnote{We note that this statistical approach has been previously 
used in Ref.~\refcite{Lima2008b} but for the SNe Ia \emph{Union} 
sample.\cite{Kowalski2008}}    

\begin{figure*}[h]
\includegraphics[scale=0.58]{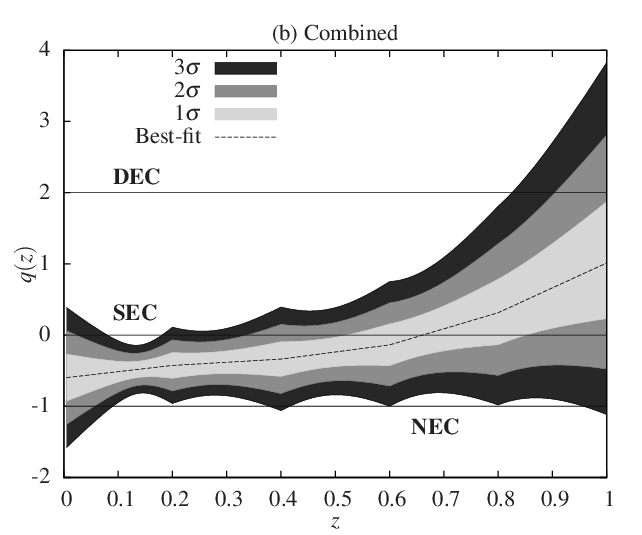}
\includegraphics[scale=0.58]{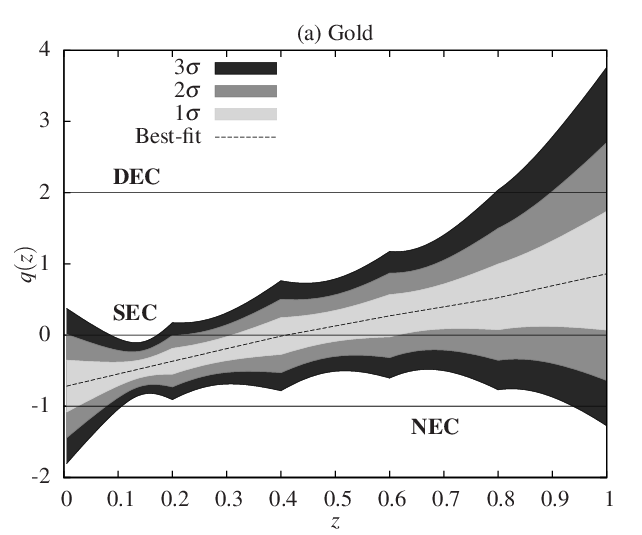}
\caption{The best-fit, the upper and lower $1\sigma$, $2\sigma$ and 
$3\sigma$ limits of $q(z_\star)$ estimates, obtained with the \emph{gold} 
[panel (a)] and the \emph{combined} [panel (b)] samples, for $200$ equally 
spaced redshifts. 
The \textbf{NEC} and \textbf{SEC} lower bounds, and also the \textbf{DEC} 
upper bound are shown. This figure shows that the \textbf{SEC} 
is violated with $1\sigma$ confidence level from $\simeq 0$ until 
$z \simeq 0.31$ [\emph{gold} sample, panel (a)], and  until $z \simeq 0.52$ 
[\emph{combined} sample, panel (b)]. It also shows that 
the \textbf{DEC} and \textbf{NEC} is violated within $3\sigma$ confidence level 
for high redshifts for both supernovae samples, and that the \textbf{NEC} is
violated for $z \lesssim 0.105$ [panel (a)] and $z \lesssim 0.085$ 
[panel (b)]. \label{qxz}}
\end{figure*}

A global picture of the breakdown and fulfillment of the energy 
conditions in the recent past has been built up with the $q(z_\star)$ 
estimates at 200 equally spaced redshifts in the interval $(0,1]$. 
Fig.~\ref{qxz}(a) shows the \textbf{NEC}, \textbf{SEC}, and \textbf{DEC} 
bounds along with the best-fit values and the $1\sigma$, $2\sigma$ and 
$3\sigma$ limits of $q(z_\star)$ in the $q(z) - z$ plane. We recall that
 \textbf{WEC} bound [$(E^2(z) \geq 0)$] is fulfilled identically.

In Fig.~\ref{qxz} it is showed that the \textbf{SEC} bound is violated 
with $1\sigma$ confidence level until $z = 0.31$ for \emph{gold} and $z = 0.52$ 
for \emph{combined} sample, while in the redshift intervals $(0.09, 0.17)$ [panel (a)] 
and $(0.08, 0.18)$ [panel (b)] this violation occurs with more than $3\sigma$
confidence level, where the highest evidence is found at $z = 0.135$
with $\simeq 3.86\sigma$ [\emph{gold}, panel (a)]  and  $\simeq 4.28\sigma$ 
[\emph{combined}, panel (b)]. 
Unlike the result of Ref.~\refcite{Lima2008a}, wherein these analyses 
have been performed by  computing the confidence regions on the 
$E(z_{\star}) - q(z_{\star})$ plane, revealing no redshift value 
for the \textbf{SEC} fulfillment with at least $1\sigma$, we note 
here the \textbf{SEC} is fulfilled with more than $1\sigma$ 
for $z \gtrsim 0.615$ [\emph{gold}, panel (a)] and $z \gtrsim 0.855$ 
[\emph{combined}, panel (b)].  
According to the present \textbf{SEC} analysis, with $1\sigma$ confidence level, 
the universe crosses over from a decelerated expansion phase to an 
accelerated expansion during the redshift interval $(\simeq 0.31, \simeq 0.615)$ 
for \emph{gold} and $(\simeq 0.52, \simeq 0.855)$ for \emph{combined} sample.%
\footnote{We recall that in a similar \textbf{SEC} analysis of Ref.~\refcite{Lima2008b} 
performed by using the \emph{Union} sample, the deceleration to acceleration 
transition expansion phase of the universe took place in the redshift interval 
($\simeq 0.4, \simeq 0.64\,$).}   

Regarding the \textbf{NEC}, Fig.~\ref{qxz} indicates its breakdown within 
$3\sigma$ confidence level for low redshift, $z \lesssim 0.105$ [panel (a)] 
and $z \lesssim 0.085$ [panel (b)]. For higher values of redshift, 
\textbf{NEC} is violated within $3\sigma$ at $z \gtrsim 0.94$ for \emph{gold} 
and $z \gtrsim 0.96$ for \emph{combined} sample. 

Concerning the \textbf{DEC}, Fig.~\ref{qxz} indicates that it is violated 
within $3\sigma$ for $z \gtrsim 0.795$ [\emph{gold}, panel (a)] and 
$z \gtrsim 0.83$ [\emph{combined}, panel (b)], which are intervals where 
the error in the estimates of $q(z)$ grow significantly, though. 
Finally, we note that the \textbf{DEC} violation of the present 
analysis is weaker than that obtained in Ref.~\refcite{Lima2008a} 
in the sense that, differently from that analysis, now the \textbf{DEC} 
is fulfilled with $1\sigma$ confidence level in the entire redshift 
interval for both samples.

\section*{Acknowledgments}

This work is supported by Conselho Nacional de Desenvolvimento 
Cient\'{\i}fico e Tecnol\'{o}gico (CNPq) - Brasil, under grant 
No. 472436/2007-4. M.P.L., S.V. and M.J.R. thank CNPq for the 
grants under which this work was carried out. We also thank A.F.F. 
Teixeira for the indication of the relevant misprints and omissions.

\end{document}